\documentclass[prl,aps,twocolumn,notitlepage]{revtex4-1}

\usepackage{stmaryrd}
\usepackage{color}
\usepackage{mathrsfs}
\usepackage{rotating}
\usepackage{graphicx} 
\usepackage{amsfonts}
\usepackage{mathrsfs}
\usepackage[american]{babel}
\usepackage{pstricks}
\usepackage{pst-plot}
\usepackage{pst-grad}
\usepackage{pst-eps}
\usepackage{slashed}

\newcommand{\bea}{\begin{eqnarray}}
\newcommand{\eea}{\end{eqnarray}}
\newcommand{\beq}{\begin{equation}}
\newcommand{\eeq}{\end{equation}}

\begin{document}
\title{Strongly Interacting Fermions and Phases of the Casimir Effect}
\author{Antonino Flachi}
\affiliation{Centro Multidisciplinar de Astrof\'{\i}sica,
Departamento de F\'{\i}sica, Instituto Superior T\'ecnico,
Universidade T\'{e}cnica de Lisboa - UTL,\\ Av. Rovisco Pais 1, 1049-001
Lisboa, Portugal}

\begin{abstract}
With the intent of exploring how the interplay between boundary effects and chiral symmetry breaking may alter the thermodynamical behavior of a system of strongly interacting fermions, we study the Casimir effect for the setup of two parallel layers using a four-fermion effective field theory at zero density. 
This system reveals a number of interesting features. While for infinitely large separation (no boundaries), chiral symmetry is broken or restored via a second order phase transition, in the opposite case of small (and, in general, finite) separation the transition becomes first order, rendering effects of finite size, for the present setup, similar to those of a chemical potential. Appropriately moving on the separation-temperature plane, it is possible to generate a peculiar behavior in the temperature dependence of the thermodynamic potential and of the condensate, compensating thermal with geometrical variations. A behavior similar to what we find here has been predicted to occur in bilayer graphene. Chiral symmetry breaking induces different phases (massless and massive) in the Casimir force separated by critical lines. 
\end{abstract}
\maketitle

It is well known that the presence of boundaries in empty space causes deformations of the quantum va\-cuum producing a macroscopic force. Casimir was the first to notice this for the case of two parallel, perfectly conducting plates and for the electromagnetic field, showing that the force, $F_c$, is attractive. In natural units, $F_c= - \pi^2 S a^{-4}/240$, with $a$ being the plate separation and $S$ their surface area \cite{casimir}.

Since its original discovery, the study of the Casimir effect has branched out in several directions and motivated experimental and theoretical research in an attempt to uncover its fundamental importance \cite{exp,books}. A somewhat special attention has been directed towards understanding how the magnitude and, especially, the sign of the force depend on the properties of the vacuum, the geometry of the boundaries, or the external conditions, and the possible technological implications of this are by now well appreciated. Regarding this point, the fermion Casimir effect has triggered some initial curiosity due to the different statistics obeyed by fermions as opposed to bosons, but it was soon understood that this difference does not necessarily lead to a change in the sign of the force. 
The application of the fermion Casimir effect that is most celebrated is, perhaps, the bag model of hadrons (see Ref.~\cite{johnson} for a review), but the scrutiny of fermion quantum vacuum fluctuations has included the study of geometrical, thermal, dimensional effects, and others (see, for example, Refs.~\cite{milton,ravndall,erdas,zanzi,jaffe,teo,fosco,bordag,elizalde,saharian,oikonomou} or Ref.~\cite{books} for a longer list of refe\-rences).

All previous work in relation to the fermion Casimir effect has focused, as far as we are aware, on the case of free fields. However, the problem may become more interesting if interactions are switched on. 
In fact, while in the case of free fields the dynamics is somewhat trivial, the inclusion of interactions opens up onto a richer spectrum of possibilities.
This problem is worthy of attention from different perspectives. 
First of all, the interplay between chiral symmetry, the geometrical and topological properties of the spacetime, and quantum vacuum fluctuations is a concrete example allowing one to explore how the phenomena of symmetry breaking and the Casimir effect are connected (see, for example, Refs.~\cite{onofrio,toms1,toms2,ford,hosotani}). Secondly, strongly interacting fermions have a well recognized importance in describing several nonrelativistic condensed matter systems routinely stu\-died in the laboratory. These include superconductors, conductive polymers, and several carbon-based materials (see Ref.~\cite{thies} for some examples). Also, since chiral symmetry resides in the quark sector, strongly interacting fermion field theories are central tools to describe the phase diagram of QCD (see Refs.~\cite{fukushima,kunihiro} for review). In this context, quantum vacuum energy effects have a direct relevance for any analysis of the phase structure aiming at including effects of finite size (see, for instance, Ref. \cite{klimenko}). 

Let us begin our discussion summarizing a few points regarding the Casimir effect for free fermions. In the following, we will consider the case of parallel plates. Natural units will be used. The free fermion dynamics is described by the Dirac equation 
with the fields forced to obey some boundary conditions at the plates that we assume to be located at $z=0$ and $z=a$.
Imposing, for simplicity, bag boundary conditions, expressed as $\left.\left(1+i\gamma^z\right)\psi \right|_{z=0,a} =0$,
leads to an implicit constraint for the momenta in the $z$ direction: 
$\Phi(k_z):= m\,{\sin(k_z a)} +k_z \cos(k_z a)=0$, where $m$ is the mass of the fermions \cite{books}.
The regularized Casimir energy, after integration over the unconstrained directions, can be written as
\bea
\mathscr{E} = -\lim_{s\rightarrow 0} {\mu^{2s}\over 2\pi} {\Gamma(s-3/2)\over \Gamma(s-1/2)}
\sum_{k_z} \left(k_z^2+m^2\right)^{3/2-s}~,
\label{cas1}
\eea
where $s$ is a regulator, $\mu$ is a renormalization scale, and the sum is over the roots of $\Phi(k_z)=0$. In the massless case, $m=0$, one finds $\mathscr{E} = -{7 \pi^2 a^{-3} / 2880 }$. When the fermion mass is nonzero, the above expression (\ref{cas1}) for the Casimir energy can be recast in the following form, 
\bea
\mathscr{E} &=& -{1\over a^3\pi^2} \int_{0}^\infty du u^{1/2}(u+\xi)(u+2\xi)^{1/2}\times\nonumber\\
&&\times \ln \left({1+{u\over u+2\xi}e^{-2(u+\xi)}}\right)+ \cdots~,
\label{eci}
\eea
where we have introduced the dimensionless quantity $\xi=m a$. The dots represent 
terms that do not contribute to the force. For $\xi=0$, expression (\ref{eci}) reproduces the massless result given above. When $\xi \neq 0$, the integral in (\ref{eci}) can be evaluated by expanding the integrand in the region of interest. For $\xi \gg 1$, 
the force is exponentially suppressed. 
Basically, $\xi$ works as a modulating parameter for the Casimir energy or force. For free fields, however, modulations caused by a change in the mass cannot occur dynamically (neither at zero nor at finite temperature or density) since, in the absence of interactions, symmetry breaking does not occur, and the mass is set by the chiral symmetry at the level of the Lagrangian \cite{ravndall}. On the other hand, when fermions are strongly interacting, chiral symmetry breaking occurs dynamically. This generates a mass for the fermions and is expected to induce a phase transition in the Casimir force. In a Casimir-like setup, the problem becomes particularly amusing, since chiral symmetry breaking can be triggered not only by thermodynamical effects, but also by changes in the geometry or topology of the system. 

Such a mechanism is reminiscent of the mass generation phenomena for self-interacting scalars in topologically nontrivial spacetimes discussed, for instance, in Refs.~\cite{toms1,toms2,ford}. There, it was a combination of scalar self-interactions and nontriviality of the geometry responsible for the spontaneous symmetry breaking. Here, the mechanism is driven by the breakdown of chiral symmetry due to thermal and geometrical effects, and it is controlled by the appearance of a condensate.

To illustrate this idea with a concrete computation, we will consider the following theory:
\bea
\mathscr{L} = 
\bar\psi i\gamma_\mu\partial^\mu \psi + {g \over 2N} \left(\bar\psi \psi\right)^2.
\label{act}
\eea
At tree level, fermions are massless and the action is invariant under discrete chiral symmetry, $\psi \rightarrow \gamma^5 \psi$. $N$ represents the number of fermion degrees of freedom and $g$ is the coupling constant. 
In the following we will use the path integral approach and introduce finite temperature effects by means of the Matsubara formalism, $t\rightarrow \imath \tau$ with $\tau \in \mathbb{S}^1$ of period $\beta=2\pi/T$, where $T$ is the temperature and antiperiodicity conditions imposed on the fermions.
The presence of the condensate, $\sigma= - g \langle \bar\psi\psi\rangle/N$, can be made explicit in the partition function, $Z$, using the Hubbard-Stratonovich transformation,
\bea
Z =\int \mathscr{D}[\bar\psi,\,\psi, \,\sigma]\, \exp
&&\left(
i \int d\tau dv_D \mathscr{L}_{\small{\mbox{eff}}}
\right),\nonumber
\eea
where $\mathscr{L}_{\small{\mbox{eff}}} = \bar\psi i\gamma_\mu\partial^\mu \psi -{N\over 2g} \sigma^2 -\sigma \bar\psi\psi$. 
In the large-$N$ approximation, an expansion of the path integral gives, 
\bea
Z = -
\int dv_D {{\sigma}^2\over 2g} + \ln \det \left(i\gamma^\mu \partial_\mu - {\sigma}\right) + O(1/N).
\label{fctdet}
\eea
The above functional determinant has to be computed consistently with the boundary conditions imposed at the plates. Approximate analytic expressions can be obtained in specific regimes of temperature and separation \cite{flachi2}. Here, we prefer to use a fully numerical approach that has the advantage of being more expedient and readily applicable to all parameter ranges. This adopts the method described, for example, in Refs.~\cite{kirsten,kirsten2,elizaldebook} to construct an appropriate contour integral representation for the functional determinant in (\ref{fctdet}), that we evaluate numerically after a convenient contour deformation. Divergences are dealt with by means of zeta function regularization, and finite temperature summations are carried out numerically.
The theory (\ref{act}) in four dimensions is nonrenormalizable and requires the introduction of a cutoff scale. Here, we will fix the coupling constant and the renormalization scale to achieve a broken symmetry phase at zero temperature in the absence of boundaries. Since the effective potential rescales as $\Omega \rightarrow \lambda^{D+1}\Omega$ under mass redefinition, $mass \rightarrow \lambda~mass$, (see Sec.~III of Ref.~\cite{flachi2}), we may fix, without loosing generality, the value of the renormalization scale (numerical value is set to $\mu=100$) and express all quantities accordingly.

\vspace{-0.3cm}
\begin{figure}[ht]
\unitlength=1.1mm
\begin{picture}(90,24)
   \put(-1.5,9){\textcolor{black}{\rotatebox{90}{$\Omega-\Omega_0$}}}
   \put(39.5,9){\textcolor{black}{\rotatebox{90}{$\Omega-\Omega_0$}}}
   \put(19,-2.6){\textcolor{black}{$\sigma$}}
   \put(59,-2.6){\textcolor{black}{$\sigma$}}
  \hskip .1cm
   \includegraphics[height=2.5cm]{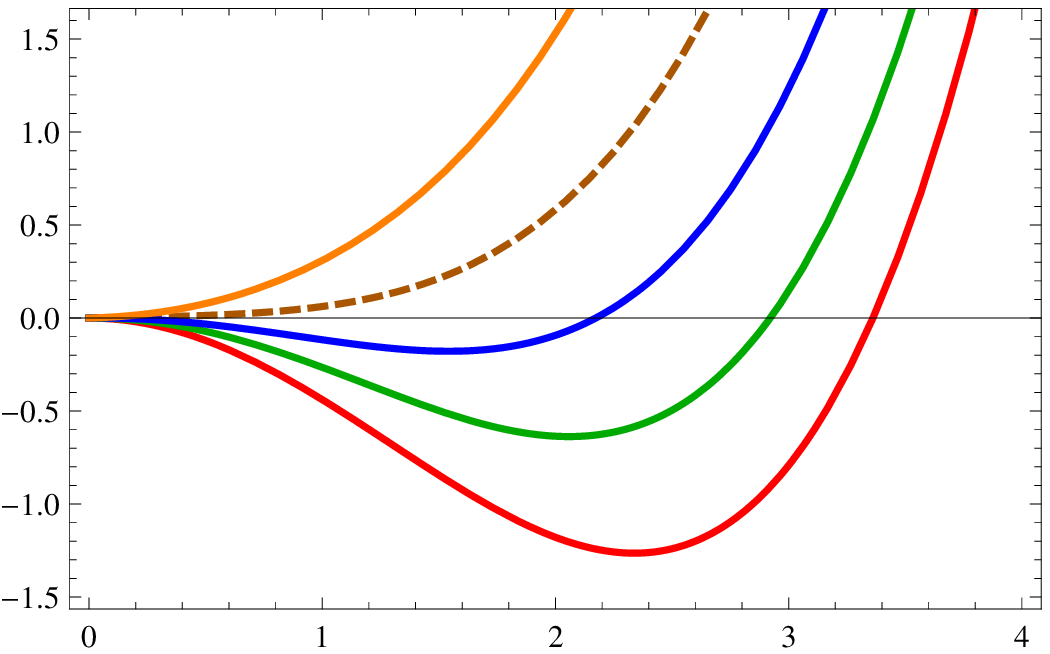}
  \hskip .5cm
  \includegraphics[height=2.5cm]{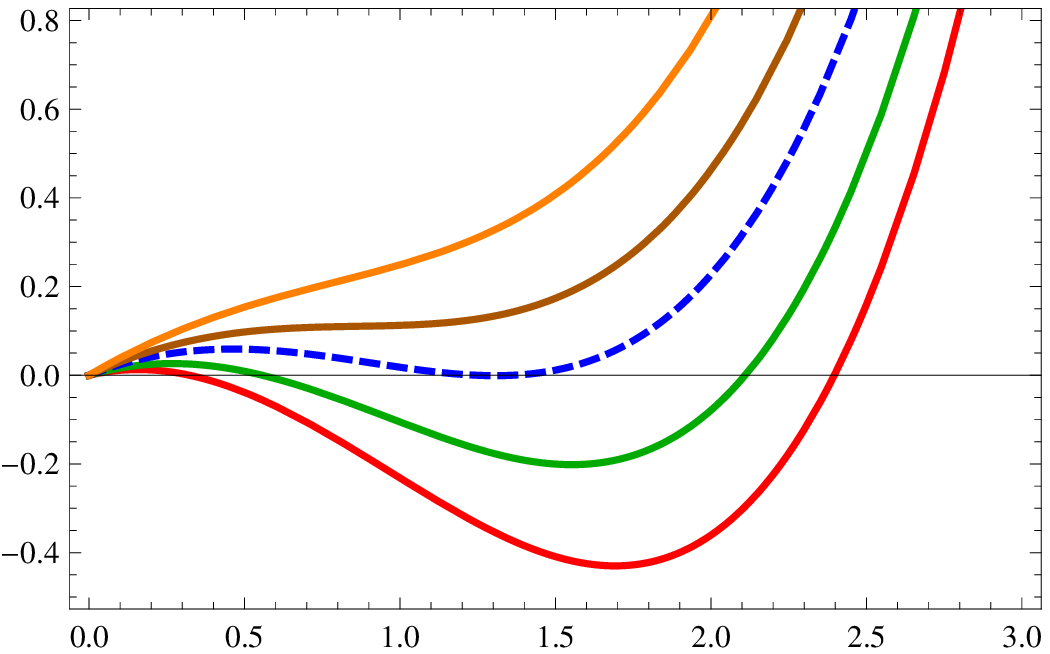}
\end{picture}
\caption{
Temperature dependence of the effective potential for fixed separation. The left-hand plot refers to the case of infinitely large separation for which the phase transition is second order. The curves refer (bottom to top) to the values of $T/T_{crit}=0.00,~0.60,~0.80,~1.00,~1.17$.
The right-hand plot refers to the case of small separation $a\times T_{crit}=0.6$
in which case boundary effects are non-negligible and the phase transition becomes first order. The curves refer (bottom to top) to $T/T_{crit}= 0.00,~0.76,~1.00,~1.13,~1.29$.
}  
\vspace{-0.3cm}
\label{fig1}
\end{figure}
\begin{figure}[ht]
\unitlength=1.1mm
\begin{picture}(90,24)
   \put(-1.5,9){\textcolor{black}{\rotatebox{90}{$\Omega-\Omega_0$}}}
   \put(39.5,9){\textcolor{black}{\rotatebox{90}{$\Omega-\Omega_0$}}}
   \put(20,-2.6){\textcolor{black}{$\sigma$}}
   \put(59,-2.6){\textcolor{black}{$\sigma$}}
  \hskip .1cm
   \includegraphics[height=2.5cm]{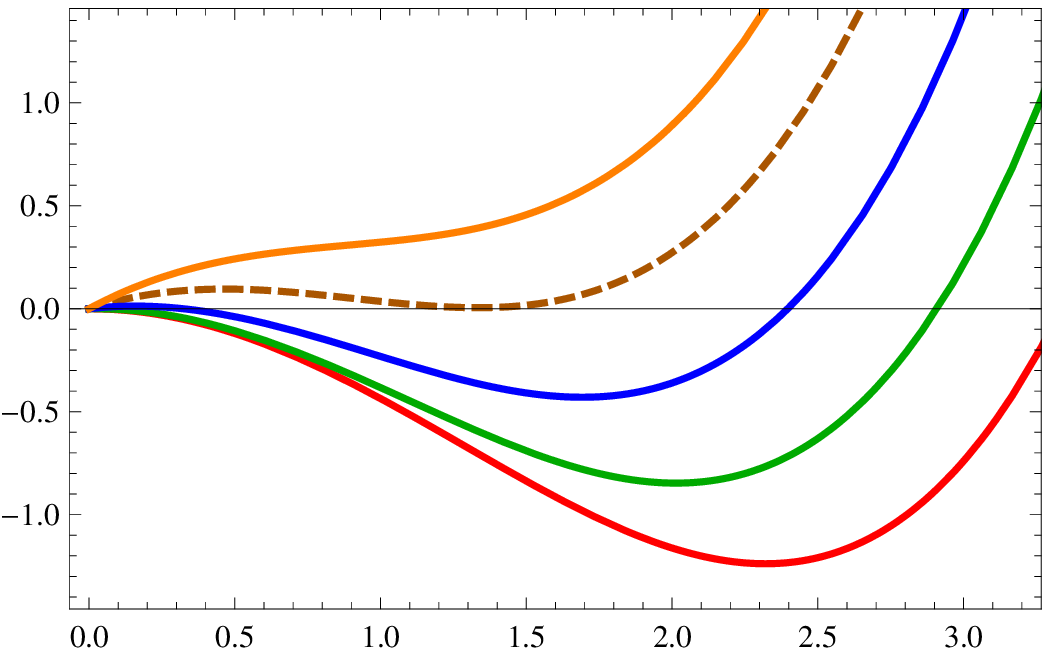}
  \hskip .5cm
   \includegraphics[height=2.5cm]{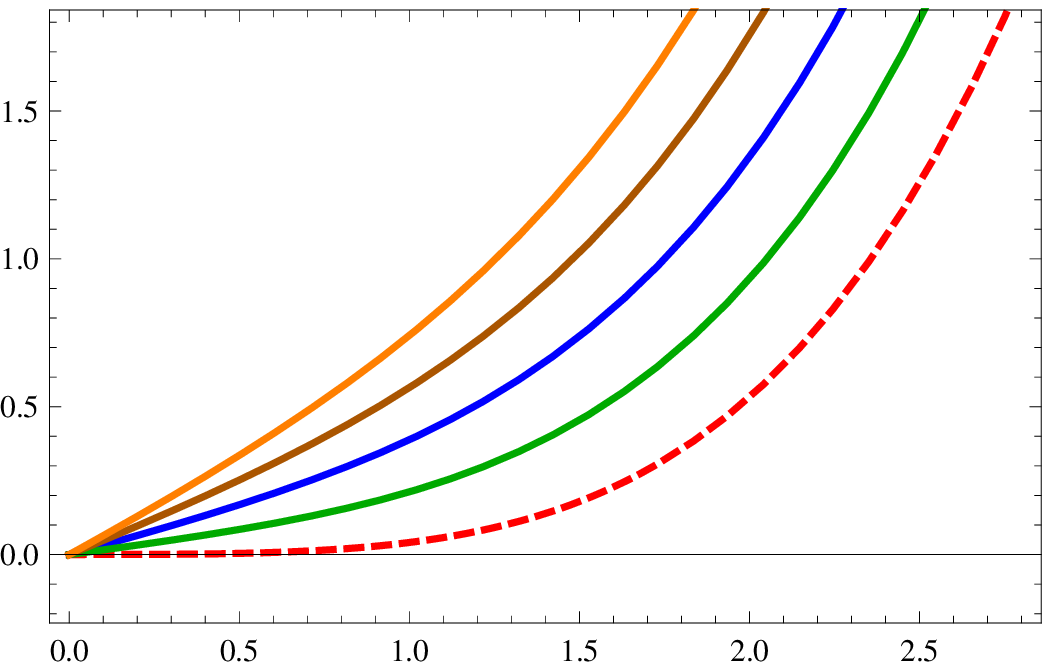}
\end{picture}
\caption{
Separation dependence of the effective potential at fixed temperature. The left-hand plot refers to the case of small temperature ($a_{crit}\times T=0.003$) and the phase transition is first order. The curves (bottom to top) refer to the following values: $a/a_{crit} = 100,~2.79,~1.39,~1.00,~0.83$. 
In the right-hand plot we have set the temperature close to its critical value in the absence of boundaries and the curves refer (bottom to top) to $a\times T_{crit} = 247.00,~7.41,~3.70,~2.47,~1.85$.
}  
\vspace{-0.5cm}
\label{fig2}
\end{figure}
\begin{figure}[ht]
\unitlength=1.1mm
\begin{picture}(90,24)
   \put(-1.5,9){\textcolor{black}{\rotatebox{90}{$\Omega-\Omega_0$}}}
   \put(39.5,9){\textcolor{black}{\rotatebox{90}{$\Omega-\Omega_0$}}}
   \put(20,-2.6){\textcolor{black}{$\sigma$}}
   \put(59,-2.6){\textcolor{black}{$\sigma$}}
   \put(9.5,9.3){\tiny{1}}
   \put(9.5,8.){\tiny{2}}
   \put(9.5,6.7){\tiny{3}}
   \put(9.5,5.3){\tiny{4}}
   \put(9.5,3.9){\tiny{5}}
   \put(9.5,2.5){\tiny{6}}
   \put(50.8,10.2){\tiny{1}}
   \put(50.8,8.6){\tiny{2}}
   \put(50.8,6.8){\tiny{3}}
   \put(50.8,5.1){\tiny{4}}
   \put(50.8,3.4){\tiny{5}}
   \put(50.8,2.){\tiny{6}}
  \hskip .1cm
  \includegraphics[height=2.5cm]{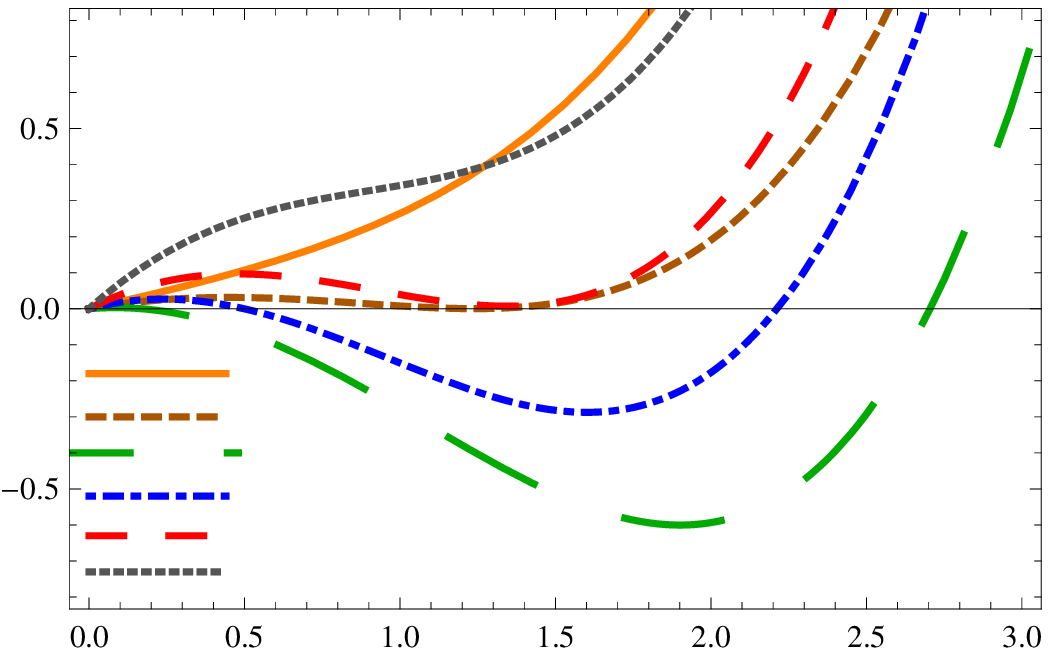}
  \hskip .5cm
  \includegraphics[height=2.5cm]{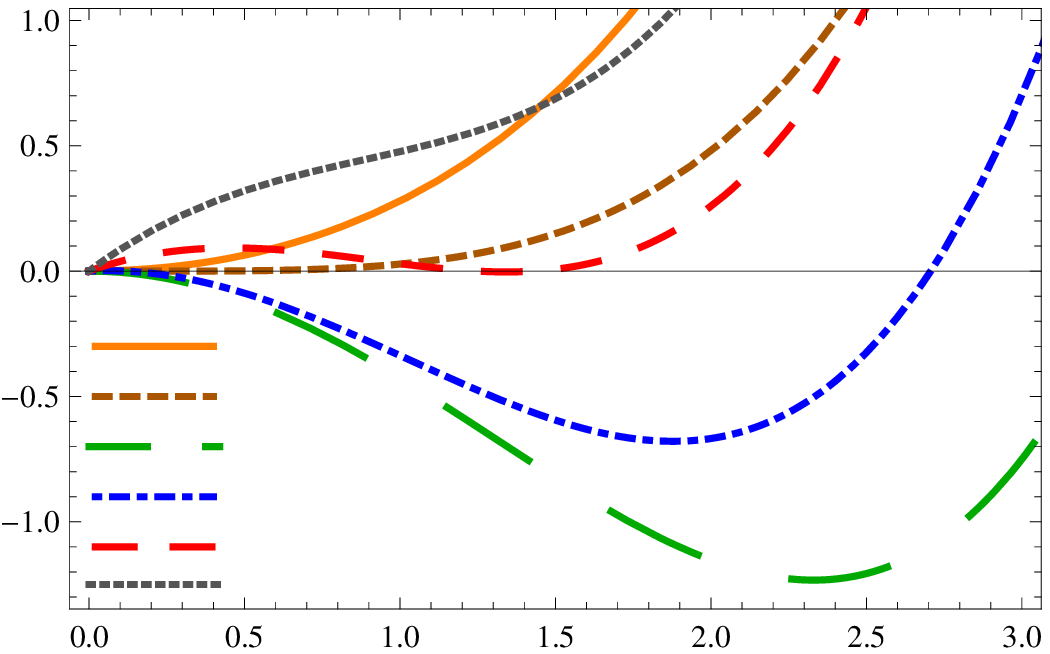}
\end{picture}
\caption{The figure shows how the effective potential changes when both temperature and separation decrease linearly according to $T(\delta)=u-v \delta$ and $a(\delta)=q-p \delta$ where $u$, $v$, $q$, $p$ are constants and $\delta$ is varied. In the left- (right-)hand panel we have set $u=q=2$ and $v=p=1$ ($u=p=1$ and $1/v=q=30$). The curves $1$-$6$ in the left- (right-)hand panel correspond, respectively, to the values of $\delta=-0.5,~0.095,~1.15,~1.55,~1.638,~1.7$ ($\delta = -60.,~-44.,~15.,~29.4,~29.64,~29.72$).}  
\vspace{-0.5cm}
\label{fig3}
\end{figure}
\begin{figure}[ht]
\unitlength=1.1mm
\begin{picture}(90,24)
   \put(-1.5,12){\textcolor{black}{\rotatebox{0}{$T$}}}
   \put(39.5,12){\textcolor{black}{\rotatebox{0}{$T$}}}
   \put(20,-2.6){\textcolor{black}{$a$}}
   \put(59,-2.6){\textcolor{black}{$a$}}
  \hskip .1cm
  \includegraphics[height=2.5cm]{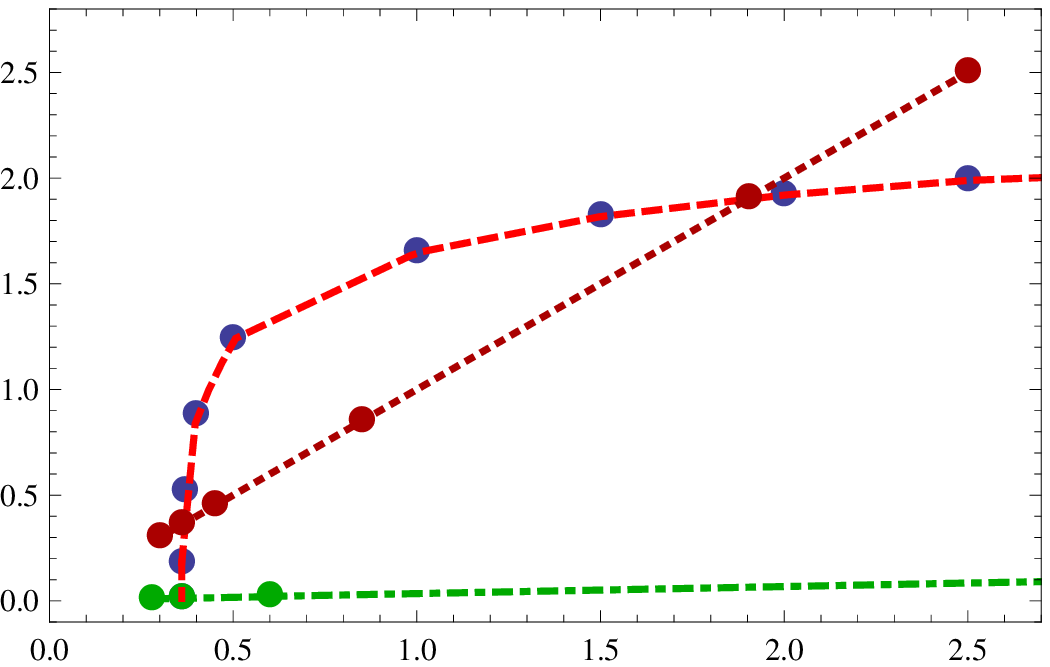}
  \hskip .5cm
  \includegraphics[height=2.5cm]{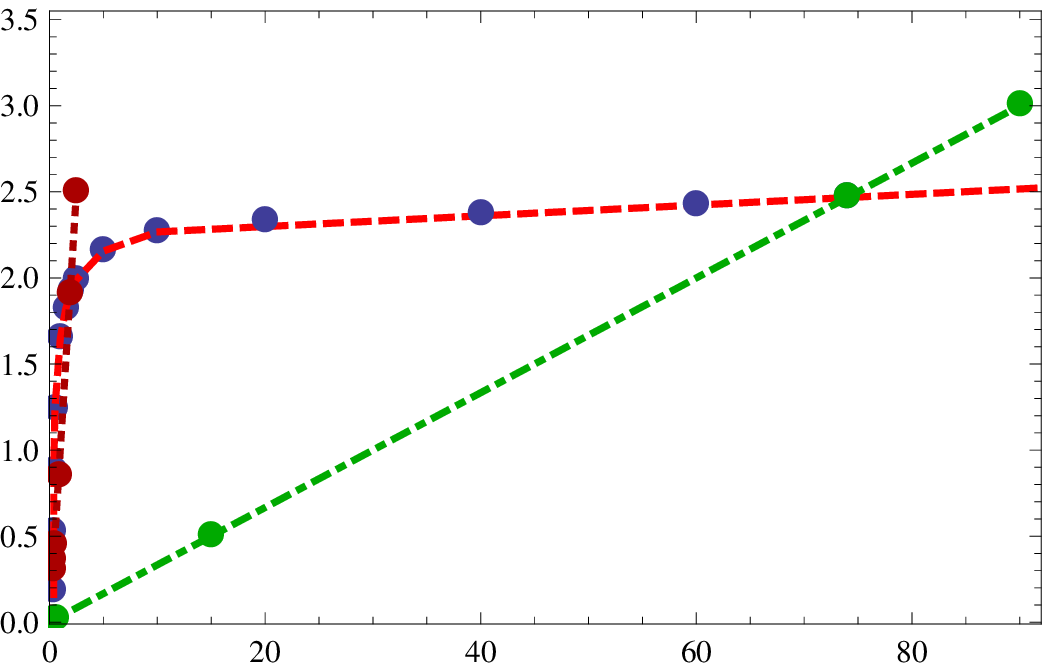}
\end{picture}
\caption{$aT$ phase diagram. The thick dashed (red) curve represents the critical line (the blue dots superposed are calculated numerically). The dots connected by the straight dotted brown (dot-dashed green) line refer to the values of $a$ and $T$ used in the left- (right-)hand panel in Fig.~\ref{fig3}.}  
\vspace{-0.6cm}
\label{fig4}
\end{figure}
\begin{figure}[ht]
\unitlength=1.1mm
\begin{picture}(90,24)
   \put(-1.5,12){\textcolor{black}{\rotatebox{90}{$\sigma$}}}
   \put(39.,9){\textcolor{black}{\rotatebox{90}{$-P_c$}}}
   \put(16.9,-2.6){\textcolor{black}{$T$}}
   \put(59,-2.6){\textcolor{black}{$a$}}
   \put(27.5,20){\tiny{0)}}
   \put(27.5,18.3){\tiny{1)}}
   \put(27.5,16.3){\tiny{2)}}
   \put(27.5,14.3){\tiny{3)}}
   \put(68,19.3){\tiny{1)}}
   \put(68,17.5){\tiny{2)}}
   \put(68,15.5){\tiny{3)}}
   \put(62,13.7){\textcolor{orange}{\tiny{massless}}}
  \hskip .1cm
   \includegraphics[height=2.5cm]{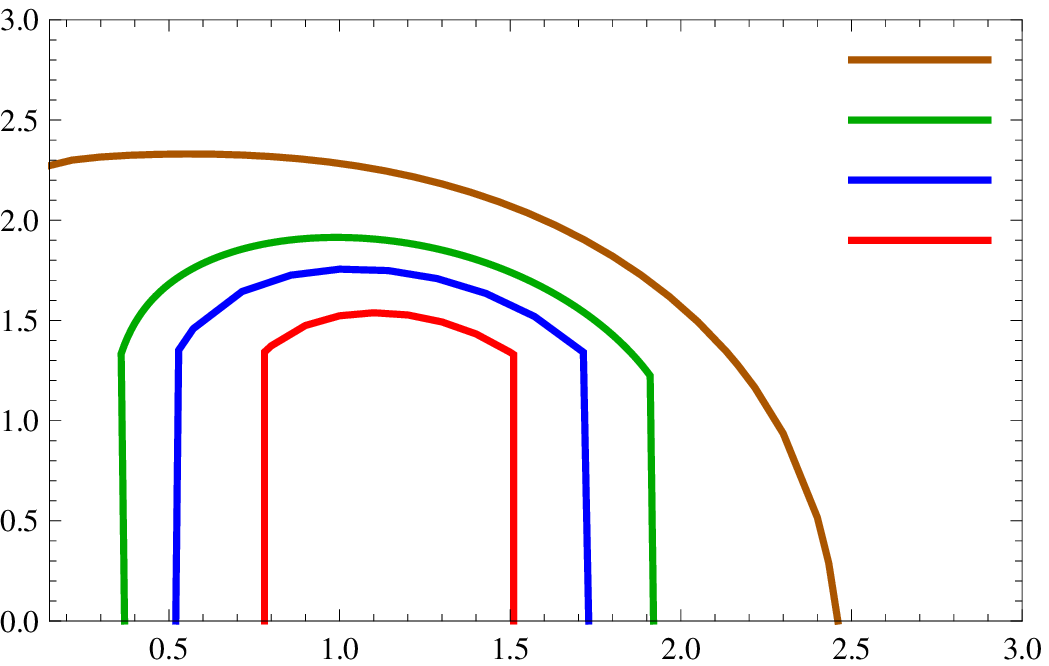}
  \hskip .5cm
  \includegraphics[height=2.5cm]{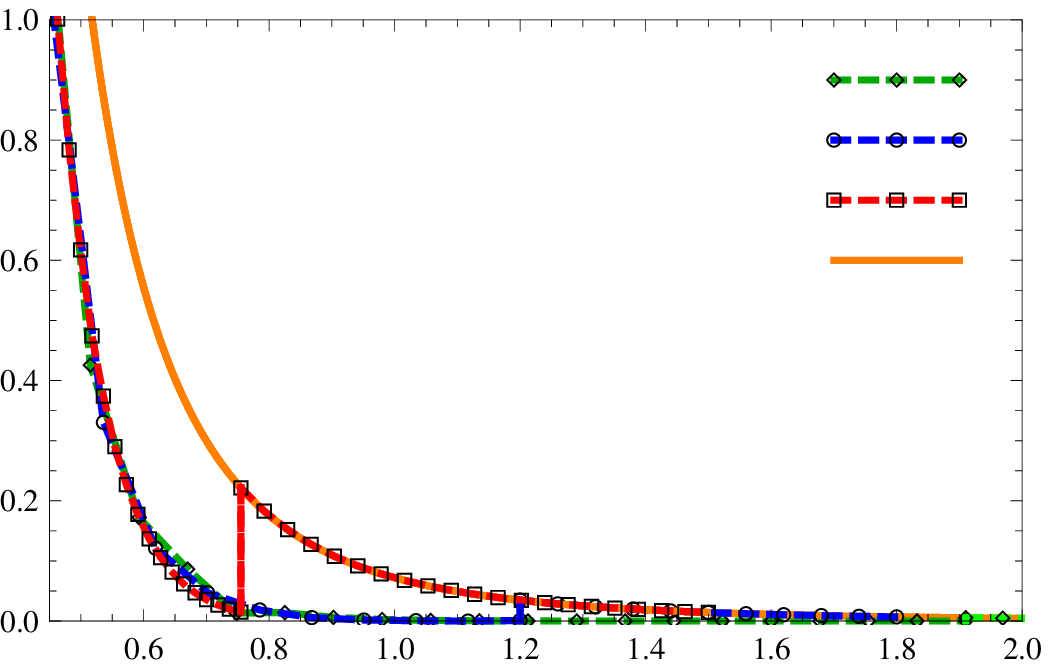}
\end{picture}
\caption{Left: Behavior of the condensate when temperature and separation change as in Fig.~\ref{fig3}. The curves refer to the following choices of the parameters: (0) $u=p=1,~q=1/v=30$; (1) $u=p=1,~q=1/v=1$; (2) $u=p=1,~q=1/v=0.7$; (3) $u=p=1,~q=1/v=0.5$. Right: Casimir pressure ($P_c=F_c/S$) for three illustrative cases. Temperature and separation change as in Fig.~\ref{fig3} with (1) $u=p=1,~q=1/v=30$; (2) $u=p=1,~q=1/v=1$; (3) $u=p=1,~q=1/v=0.7$. The orange continuous curve refers to the massless case.}  
\vspace{-0.3cm}
\label{fig5}
\end{figure}

Our results are summarized in Figs.~\ref{fig1}-\ref{fig5}. Figure \ref{fig1} shows how the effective potential $\Omega$ (normalized by subtracting its value $\Omega_0$ at $\sigma=0$) changes with temperature when the separation $a$ is fixed. The right- (left-)hand panel refers to small (large) values of the separation $a$, and chiral symmetry breaking occurs via a first (second) order phase transition. 
Figure \ref{fig2} illustrates the opposite situation, that is, the dependence of the effective potential on the separation $a$ for fixed temperature. The left-hand panel of Fig.~\ref{fig2} refers to the case of very small temperature, while in the right-hand panel we have set $T$ close to its critical value in absence of boundaries. 
From Figs.~\ref{fig1} and \ref{fig2}, it is clear that a decrease of $a$ tends to increase the effective temperature of the system. However, when the separation is small and both geometrical and thermal effects are important, the phase transition is a first order one, in contrast with a second order transition expected when changes are purely thermodynamical and boundary effects negligible. In this sense, effects of finite size for the present geometry seem to be more similar to those of a chemical potential than of temperature. 

The nature of the change in the order of the transition can be better understood using the Ginzburg-Landau expansion for the thermodynamic potential,
\bea
\Omega - \Omega_0 &=& 
c_0(a,T)\sigma^2
+ c_1(a,T)\sigma^3+ c_2(a,T)\sigma^4 +\cdots~,
\nonumber
\eea
where the coefficients $c_i(a,T)$ are dimensionful functions of the temperature and separation and depend on the geometry, topology, and the boundary conditions. In the absence of boundaries, {\it i.e.}, for $a\rightarrow \infty$, chiral symmetry, $\sigma \leftrightarrow -\sigma$,  prohibits the appearance of odd powers in the above expansion and guarantees the coefficients associated with such terms to vanish, leading to a second order phase transition. In the present case, however, dimensional inspection indicates the presence of terms proportional to $a^{-1}\sigma^3$, and explicit computation of the associated coefficient shows that they do not vanish (see Ref.~\cite{flachi2}). This causes the transition to change to first order.
A deeper connection can be drawn using the Schwinger--De Witt expansion that relates the coefficients $c_i(a,T)$ to the heat-kernel coefficients $\theta_{i/2}$ asso\-ciated with the operator in (\ref{fctdet}), $c_i(a,T) \propto \theta_{i/2}$ (see Ref.~\cite{flachitanaka}). In this way, it is possible to see that odd powers of the condensate in the Ginzburg-Landau expansion are accompanied by half-integer (boundary) heat-kernel coefficients. These (see Ref.~\cite{kirsten} for explicit expressions of $\theta_{i/2}$ and Ref.~\cite{schaden} for a clear discussion in relation to the Casimir effect) are related to the boundary geometry, topology, and also to the boundary conditions that, in the present situation, break chiral symmetry.

The above results suggest that simultaneous changes in temperature and separation may have interesting effects. In fact, if a decrease in temperature typically pushes down the potential and tends to bring the system into a broken phase with a nonvanishing condensate, a decrease in distance has the opposite effect. Therefore, by moving appropriately on the ($a$-$T$) plane, it is possible to compensate thermal with geometrical changes. This may result in a peculiar phase diagram with the system moving towards a broken chiral symmetry phase despite an increase in temperature. An example of this sort is shown in Fig.~\ref{fig3}. 
To illustrate this point, let us first look at the left-hand panel of Fig.~\ref{fig3}. The initial configuration (curve $1$) is obtained by setting $T$ large and $a$ small. In this case, the effective temperature is high and the system is in a symmetric phase. We then start decreasing both temperature and separation. (For illustration, we vary $a$ and $T$ linearly, according to the relations $T(\delta)=u-v \delta$ and $a(\delta)=q-p \delta$, where we set $u=q=2$, $v=p=1$ and change $\delta$). The initial configuration (curve $1$) refers to $\delta=-0.5$. Increasing $\delta$ ({\it i.e.} lowering temperature and separation), the effective potential moves downwards and we encounter a second order phase transition for $\delta\simeq 0.095$ (curve 2). A further increase in $\delta$ keeps pushing the potential down and the condensate up until $\sigma$ reaches a maximum (for $\delta\simeq 1.15$, curve $3$). In this process, the breakdown of chiral symmetry is driven by the temperature, assisted by effects of finite size that are responsible for the change in the order of the transition. From this configuration, a further increase in $\delta$ changes the tendency of the potential that, despite the decrease in temperature, starts to move upwards (and in $\sigma$ that descends from the maximum). When $T$ and $a$ reach a critical value (for $\delta\simeq 1.638$, curve 5), another second order transition occurs restoring chiral symmetry. In this region, it is the separation that dominates over the temperature. Starting from a configuration where the initial separation is large, it is possible to reduce the effects of the boundaries adjusting the order of the first phase transition, essentially, to a first order one. The right-hand panel of Fig.~\ref{fig3} refers to this case. 
Figure \ref{fig4} shows the phase diagram in the region of small $a$ (left-hand panel) and over a larger range of separation (right-hand panel). The dots connected by the straight lines refer to the values of $a$ and $T$ used in Fig.~\ref{fig3}. Finally, in the left-hand panel of Fig.~\ref{fig5} we illustrate how the condensate changes when both $T$ and $a$ vary linearly as in Fig.~\ref{fig3}.
\vspace{-0.3cm}

The effects of chiral symmetry breaking on the Casimir effect should now be clear. 
In essence, for a system of strongly interacting fermions, it is chiral symmetry that controls the fermion mass through the appearance of a nonvanishing condensate and, in turn, the parameter $\xi=ma$, related to the suppression factor in the Casimir force. 
With the previous results in hands, we have computed the Casimir force between the layers, and results are shown in the right-hand panel of Fig.~\ref{fig5}. The thick orange curve represents the pressure in the massless case, while dashed curves refer to the interacting case with $T$ and $a$ varied as before. Because of the temperature dependence of the condensate, we, in fact, expect two phase transitions: one at smaller and one at larger values of $a$. However, effects due to the transition at the smaller value of $a$ are negligible as it can be understood by estimating the parameter $\xi$ that is bounded by the separation times the value attained by the condensate in the infinite volume limit, {\it i.e.} $\xi < O(10^{-1})$. On the other hand, the transition occurring at the larger values of the separation leads to sizable effects. Notice, also, that when the initial separation is larger (see curve 1 in Fig.~\ref{fig5} (right)), suppression of the Casimir force occurs over all the separation range. For the case of initial smaller separation (see curve 3 in Fig.~\ref{fig5} (right)), the suppression occurs over a smaller separation range and the phase transition is clearly visible. 

We would like to note here that a behavior similar to Fig.~\ref{fig3} has been predicted to occur in bilayer graphene with interlayer pairing of electrons 
and it is expected to induce an exotic, possibly experimentally observable, form of interlayer superconductivity \cite{hosseini}. 
Whether an actual Casimir effect experiment using bilayer graphene (or other strongly coupled fermionic materials) can be performed leading to any observable effect of fermion quantum vacuum fluctuations is currently under investigation. Certainly, it seems interesting to examine whether quantum vacuum energy effects, indirectly, may affect the properties of these systems (e.g., stability, separation-temperature correlations, etc.), manifes\-ting yet a new facet of symmetry breaking, and, eventually, indicate new directions of study for the Casimir effect.

Our goal was to look at the interplay between chiral symmetry breaking and boundary effects in the tractable and nontrivial case of two parallel layers, for a system of strongly coupled fermions. A number of intriguing features arose. First, finite size effects, for the present setup tend to change the order of the phase transitions from second order (in the infinite volume limit) to first order, rendering effects of finite size for the present setup more similar to those of a chemical potential than of temperature. Whether and how an appropriate choice of geometry, topology, and boundary conditions may preserve chiral symmetry is certainly an important question that could be relevant in the context of finite temperature or density QCD. Second, we have shown that simultaneous changes in temperature and separation may induce an interesting behavior in the thermal dependence of the potential and of the condensate that resembles what has been discussed earlier for the case of bilayer graphene. This similarity provides, in our opinion, motivation to consider these issues further. Finally, we have shown how chiral symmetry directly influences the Casimir effect, inducing different phases in the force separated by critical lines providing a new example linking the phenomena of symmetry breaking with quantum vacuum energy effects.  

{\it Acknowledgements.} I gratefully acknowledge the support of the Funda\c{c}\~{a}o para a  Ci\^{e}ncia e a Tecnologia of Portugal and of the Marie Curie Action COFUND of the European Union Seventh Framework Programme (grant agreement PCOFUND-GA-2009-246542). Thanks are extended to D. Antonov, Y. Gusev, J. Lemos, J. Rocha, and T. Tanaka for discussions and to the Referees for the careful scrutiny of the manuscript and important suggestions. I also wish to express my gratitude to the Yukawa Institute for Theoretical Physics of Kyoto University for the kind hospitality during the long-term workshop `Gravity and Cosmology' (YITP-T-12-03). 

\end{document}